\documentclass{ws-ijmpd}

\begin{document}

\markboth{R. Opher and A. Pelinson} {A New Test for Dark Energy
Models}

%
\catchline{}{}{}{}{}
%

\title{A New Test for Dark Energy Models}

\author{Reuven Opher}

\address{Instituto  de Astronomia, Geof\'{\i}sica e Ci\^encias Atmosf\'ericas,
Universidade de S\~{a}o Paulo\\
Rua do Mat\~{a}o 1226, Cidade Universit\'aria, CEP 05508-900, S\~{a}o Paulo, S.P., Brazil\\
opher@astro.iag.usp.br}

\author{Ana Pelinson}

\address{High Energy Physics Group, Dept. ECM, Universitat de Barcelona\\
Diagonal 647, E-08028, Barcelona, Catalonia, Spain\\
apelinson@ecm.ub.es}

\maketitle

\begin{history}
\received{Day Month Year} \revised{Day Month Year} \comby{Managing
Editor}
\end{history}

\begin{abstract}
One of the greatest challenges in cosmology today is to determine
the nature of dark energy (DE), the source of the observed present
acceleration of the Universe. Besides the vacuum energy, various
DE models have been suggested. The tests that have been proposed
to differentiate among these models are based on observations of
galaxies at high redshift ($z>0$), to be obtained in the future.
We suggest here a new test that is valid at $z\simeq 0$. It is
based on existing observational data, numerical simulations, and
three well known analytic models that evaluate the bias parameter
$b$, the ratio of galaxy to dark matter (DM) fluctuations. These
analytic models are based on the physical processes involved in
the formation of stars and in the formation and merging of
galaxies. The value of $b(z)$ obtained in each model is a function
of the DM growth factor $D(z)$, which, in turn, is a function of
the DE. We show that the equations for $b$ in all three analytic
models can be reduced to the form of a known constant plus the
term $E[D(z=0)/D(z)]^{\alpha}$, where $\alpha=1$ or 2 and $E$ is a
free parameter. Using the value of $b$ obtained by the 2dFGRS
consortium for the $\Lambda$CDM model, to normalize $E$, we find
that all three analytic models predict $b^2(0)=1\pm 0.1$ for all
DE models. Since we use the result that $b^2(0)\simeq 1$ from the
2dFGRS consortium for the $\Lambda$CDM model, the $L_\ast$ galaxy
used in our text is the same as that obtained by the consortium
for a broad range of galaxy types, using the Schechter function
fit to the overall luminosity function of their $\sim 220,000$
galaxies. Numerical simulations that evaluated $b^2(0)$ for the
$\Lambda$CDM and CDM ($\Lambda=0$) models also obtained
$b^2(0)=1\pm 0.1$. Since this value of $b^2(0)$ is indicated by
numerical simulations as well as by all three popular analytic
models, which are normalized by the 2dFGRS consortium result for
the $\Lambda$CDM model, we suggest the condition that $b^2(0)=1\pm
0.1$ at $z=0$ as a new test for the viability of DE models.  Thus,
for a given observed galaxy fluctuation spectrum such as that of
the 2dFGRS consortium, if the DM fluctuations are greater or less
than the galaxy fluctuations by more than $10\%$, the DE model can
be discarded. As examples of this test, we show that three popular
DE models do not satisfy this test: the vacuum metamorphosis model
deviates from $b^2(0)=1.0$ at $z=0$ by $20\%$, the brane-world
model by $26\%$ and the supergravity (SUGRA) model by $38\%$.

\end{abstract}

\section{Introduction}

The nature of dark energy (DE), the existence of which was first
indicated to explain the recent SNIa observation of the
acceleration of the Universe, is one of the major problems in
cosmology.\,\cite{Riess,Perl} Theories in which gravity is
modified as well as those that include parametrizations of the DE
equation of state (EOS), $w(z)=p/\rho$, where $p\,(\rho)$ is the
pressure (energy density) of the DE, have been suggested to
explain it.\,\cite{wag86}\cdash\cite{lindprl} Based on
observations, various constraints have been put on the EOS for a
variety of models (see e.g., Refs.
\refcite{huttur}--\refcite{comm2a}). In order to investigate DE
models, we used the bias parameter $b^2$, the ratio of galaxy
clustering $\xi_{gg}$ to dark matter (DM) clustering $\xi_{dm}.$
\par
Observed galaxy clustering $\xi_{gg}$ [or its power spectrum
$P(k)$] does not directly provide information about the DE. The DE
affects the DM clustering $\xi_{dm}$ which, in turn, affect
$\xi_{gg}.$ In order to obtain information about DE, we need to
know the ratio $b^2=\xi_{gg}/\xi_{dm}.$ In the future,
observations will be used to determine $\xi_{dm}$. Thus, along
with observations of $\xi_{gg}$, we will be able to obtain $b^2$.
One very important observational project to determine $\xi_{dm}$
is the DES (the Dark Energy Survey)
(http://www.darkenergysurvey.org). DES will study the growth of DM
fluctuations as a function of redshift with ``weak" gravitational
lensing, produced by the DM fluctuations. The results of DES will
take some time, however, to become available. In the present
paper, we use observations to normalize existing analytic models
to predict $b^2(0)\simeq 1$ at $z = 0$, which we propose as a new
test for DE models. At the present time, this test is very useful
for limiting viable DE models. Recently, Grande et al. used it to
limit the validity of two DE models.\,\cite{Grande07}

\par
Somerville et al. define the general bias parameter as
$b(\delta)\,\delta\equiv  < \delta_g|\delta
>= \int{d \delta_g}\, P(\delta_g|\delta)\, \delta_g$ for a  DM
fluctuation $\delta$.\,\cite{somerville01} The bias parameter
$b(\delta)$ is the average of the probability $
P(\delta_g|\delta)$ that there is a galaxy fluctuation $\delta_g$
within the matter fluctuation $\delta$. This relation fully
characterizes the mean non-linear biasing and reduces to the
linear biasing relation, $b\,\delta(\lambda)=\delta_g(\lambda)$,
if $b$ is independent of $\delta$, where $\delta_g(\lambda)$ is
the galaxy fluctuation in a sphere of radius $\lambda$. In order
to track the formation of galaxies and quasars in their
simulations to evaluate $b$, Somerville et al. used a
semi-analytic model to follow gas, star, and supermassive black
hole processes within the merger trees of DM halos and
substructures. This semi-analytic model is described in Refs.
\refcite{somerville99,kau00,springel01}. The modelling assumptions
and parameters were adjusted in order to fit the observed
properties of low redshift galaxies, primarily their joint
luminosity-color distribution and their distributions of
morphology, gas content, and central black hole mass.
\par
The observed galaxy power spectrum from the final 2dFGRS catalogue
can be found in Ref.\,\refcite{cole05} (Cole et al.). According to
Cole et al., the large-scale linear bias factor for $L_\ast$
galaxies, $k\le 0.1 h\,{\rm {Mpc}^{-1}}$ is $b=1.03$ for the
$\Lambda$CDM model. This is consistent with their previous result
for brighter $L_S$ galaxies, $b(L_S,z=0)=1.10\pm 0.08$, obtained
from APM-selected massive galaxies ($L_S=1.9
L_{\ast}$).\,\cite{2df}
\par
Bias depends on the luminosities of the galaxies studied and is
known as ``luminosity segregation". In order to normalize the bias
to $L_\ast$ galaxies, the 2dFGRS consortium used the expression
$b(L)=0.85+0.15 (L/L_\ast)$ to take into account luminosity
segregation.\,\cite{springel01} Since $L_{\ast}$ galaxies are
almost unbiased, the 2dFGRS consortium result indicates that
$b^2(z\cong 0)\simeq 1.0$ to a $10\%$ accuracy for the
$\Lambda$CDM model. We use this important result in our new test
for DE models.
\par
The comparison between the DM fluctuations at the CMB
epoch and those at the present, involves two steps:\\
\noindent(i) the calculation of the DM perturbation amplitude at
the CMB last scattering epoch $\xi_{dmls}$ from the observed
$C_l$ of the CMB and the cosmological parameters; and\\
(ii) the calculation of the DM perturbation amplitude at
the present epoch $\xi_{dm0}$ from the growth rate and
$\xi_{dmls}$.
\par
We assume negligible DE at the
recombination era in the present paper, making the perturbation amplitude at
recombination independent of DE models. This is a
reasonable assumption since the flatness of the
Universe, which we assume for all our DE models, primarily
determines the sound horizon for the CMB data and the mapping
between the matter power spectrum $k$ and the CMB $l's$
[step (i)],while the DE primarily determines the growth rate [step (ii)].
\par
The perturbation amplitude at the recombination era depends
primarily on the cosmological parameters $\Omega_M,\,\Omega_B$,
and $H_0$ in a flat Universe. The parameter $\Omega_M$ is generally
attributed to neutralinos produced in the supersymmetric dominated
primordial Universe, $\Omega_B$ is produced in the
baryogenesis primordial Universe, and the present Hubble parameter
$H_0$ in a flat Universe is due to $\Omega_M^0$ and $\Omega_B^0$.
\par
These parameters can be obtained from observations, using methods
that are not in any way at all connected with the CMB and DE
models. The $\Omega_M$ can be obtained from high precision
observations of galaxy clusters, using the relation
$\Omega_M=(M/L) J/\rho_{c}$, where $M(L)$ is the mass (luminosity)
of a given cluster, $J$ is the luminosity density in the field
around the cluster, and $\rho_{c}$ is the critical density. Other
methods using clusters and galaxies that depend on the linear
theory growth factor (and thus, on the DE model), are also used to
determine $\Omega_M.$ The parameter $\Omega_M$ can be obtained
from the observed number of clusters as a function of redshift,
compared with the number predicted by the Press-Schechter relation
or the Sheth-Tormen model,\,\cite{Sheth} which depends on the
linear theory growth factor and the DE model. Yet another method
of obtaining $\Omega_M,$ which depends on the linear theory growth
factor, uses $\sigma_8,$ the present average amplitude of the dark
matter fluctuations in a sphere of radius $8 h^{-1}$Mpc. The
parameter $\Omega_B$ can be determined from high precision
deuterium abundance observations of quasar absorption lines, and
the present parameter $H_0,$ from data of Cepheids in nearby
galaxies.
\par
Analytic models that evaluate the linear bias evolution $b(z)$ as
a function of redshift, depend on the EOS of the
DE.\,\cite{Munshi} These analytic models predict $b^2(z\sim
0)\simeq 1.0$ for the flat DE $\Lambda$CDM model. The reason for
this was noted by Tegmark and Peebles:\,\cite{tegmark98} ``...even
if galaxies initially were uncorrelated with the mass, they would
gradually become correlated as gravity draws them toward overdense
regions and one might expect this process to drive $b$ toward
unity". Numerical simulations, such as those of Somerville et al.,
also obtain this value for $b^2(0)$ for very different DE models,
for example, the CDM ($\Lambda=0$) and $\Lambda$CDM models.
\par
From the analysis of the 2dFGRS data by Cole et al., $b^2_{\Lambda
CDM}(0)\simeq 1.06$. Normalizing the analytic bias models,
discussed in Section~2, to $0.9\leq b^2_{\Lambda CDM}(0)\leq 1.1$,
we find that $0.9\leq b^2(0)\leq 1.1$ for viable DE models, which
we propose as a new test for DE models. Since we use the result
that $b^2(0)\simeq 1$ from the 2dFGRS consortium for the
$\Lambda$CDM model, the $L_\ast$ galaxy used in our text is the
same as that obtained by the consortium for a broad range of
galaxy types, using the Schechter function fit to the overall
luminosity function of their $\sim 220,000$ galaxies.
\par
We apply our DE model test to several popular DE
models. We relate the $b^2$ of the DE model at $z\sim 0$
to a factor, $F=|b^2-b_\Lambda^2)/b^2_\Lambda|,$ where $b_\Lambda^2$ is the $b^2$ of the
flat $\Lambda$CDM model at $z\sim 0$. A $10\%$ deviation of $b^2$
from $b_\Lambda^2$ normalized to unity implies a $10\%$ deviation of $F$
from zero, $F=0.1$. Viable DE models then need to have $F
\leq 0.1$ (i.e., $b^2=1.0$ to a $10\%$ accuracy at $z=0$).
\par
Three well-known analytic models of linear bias evolution are
discussed in Section~2. We discuss the effect of DE on the linear
growth of $(\delta\rho/\rho)$ in Section~3. DE models can be
described by an EOS, $P/\rho=w(a)=w_0+w_a(1-a)$, where $P(\rho)$
is the pressure (energy density) at the cosmic scale factor
$a$.\,\cite{comm1} It is shown in this section, that the
permissable values of $F\leq 0.1$ limits the parameters $w_0$ and
$w_a$. We then go on to discuss three popular DE models: the
five-dimensional brane-world model (BWM),\,\cite{bwm} the vacuum
metamorphosis model (VMM),\,\cite{vmm} and the supergravity
(SUGRA) model.\,\cite{sugra} Conclusions and discussion are
presented in Section~4.

\section{Analytic bias models as a function of DE}

As noted above, the 2dFGRS final results as well as the numerical
calculations of Cole et al. indicate a bias parameter $b(z\cong
0)\simeq 1.03$ for the $\Lambda$CDM model. From these results, we
have $b^2_{\Lambda CDM}\simeq 1.0$ to better than a $10\%$ accuracy.
We use this result to normalize well-known analytic models for
$b^2(z)$ in order to obtain $b^2(0)$ for viable DE models.

\subsection{Galaxy Conserving Model (GCM)}

In this model, galaxies behave as test particles, with their
intrinsic properties conserved (see e.g., Refs.
\refcite{tegmark98,Nusser,Matarrese}). An $L_\ast$ galaxy is a
present massive galaxy. It probably started to form as a small
galaxy at a high redshift, gradually building up by the process of
merging. The assumption of the creation of an $L_\ast$ galaxy at a
high redshift, with no further creation or merging, is an
approximation of the evolution of $L_\ast$ galaxies. The linear
bias parameter relating the density distribution to its mass
density for a galaxy population formed at a given cosmic epoch
$z_\ast$ is given by
\begin{equation}
b_{GCM}(z)=1+ [b(z_\ast) -1]\frac{D(z_\ast)}{D(z)}=1+ (b_0
-1)\frac{D(z=0)}{D(z)}\,,
\end{equation}
where $b_0$ is the bias parameter at the present epoch and $D(z)=
\delta\rho/\rho$ is the linear growth of the density fluctuations
(details of $D(z)$ are presented in the Section~3).

\subsection{Galaxy Merging Model (GMM)}

The evolution of galaxy clustering is associated with host dark
matter halos in this model. We use the analytical expression
obtained by Ref. \refcite{Mo} for the halo-halo correlation,
\begin{equation}
\xi_{hh}(r,M)= b^2(M)\, \xi_{mm}(r)\,,
\end{equation}
where the bias parameter, according to the Press-Schechter
formalism,\,\cite{Press} can be written as
\begin{equation}
b_{GMM}(M,z) = (1-\frac{1}{\delta_c}) + (\nu^2/\delta_c)\,
\frac{D^2(z=0)}{D^2(z)}\, \label{bmerg},
\end{equation}
where $\nu$ is the ratio of $\delta_c$ to the average value
of the DM fluctuation $\sigma(M),$ for the mass $M,$ [Eq.(5)].
From the dynamics of the spherical collapse in an expanding
background, the factor $\delta_c$ was derived, and shown to be
$\delta_c\simeq 1.69$, independent of the DE
 model by Ref.
\refcite{Weinberg}. The factor
\begin{equation}
\nu^2/\delta_c={\delta_c}/{\sigma^2(M)}\, \label{Edef}
\end{equation}
is independent of $z$ and $\sigma^2(M)$ is defined as
\begin{equation}
{\sigma^2(M)}=\frac{\sigma^2(M,z)}{D^2(z)/D^2(z=0)}\,,
\end{equation}
where
\begin{equation}
\sigma^2(M,z)=\frac{D^2(z)/D^2(z=0)}{2\pi^2}\int_0^{\infty}{k^2}
P(k)W^2(k,M) dk \,
 \label{sigma}
\end{equation}
is the rms of the linear density fluctuation of top-hat spheres
containing an average mass M. The function $P(k)$ is the linear
power spectrum at redshift zero and $W(k,M)$ is the Fourier space
representation of the real space top-hat enclosing the average
mass M.

\subsection{Star Forming Model (SFM)}

DM halos with masses greater than a given mass $M$ can be
identified with galaxies with luminosities greater than a
corresponding luminosity $L$ at a redshift $z$ in the star-forming
model (see e.g., Ref. \refcite{Blanton}). In this model, the bias
evolution is
\begin{equation}
b_{SFM}(M,z)=1+ \,(\nu^2/\delta_c)\, \frac{D^2(z=0)}{D^2(z)}\,.
\end{equation}
\par
The equations for $b(z)$ in the above three models (i.e., Eqs.~(1),
(3) and (7)) can all be written in the form
\begin{equation}
b(z)=A + E \left[\frac{D(z=0)}{D(z)}\right]^{\alpha}\,,
\end{equation}
where $A$ is a known constant, $\alpha=1$ or 2, and $E$ is a free
parameter. For the $\Lambda$CDM model obtained by the 2dFGRS
consortium, $b^2=1.0\pm 0.1$ at a mean redshift
$z=0.17$.\,\cite{springel01} The curves for the star forming and
galaxy conserving models in Fig.~1 were normalized to
$b^2(0)=1.1,$ the approximate maximum, value of the 2dFGRS
consortium. From Eqs.~(1), (7) and Fig.~1 we note that the galaxy
conserving model has $b^2$ equal to a constant plus a term
$\propto z$ and the star forming model a constant plus a term
$\propto z^2$. Both models predict a change of $b^2$ from $z=0.17$
to $z=0$ by a very small factor $\sim 0.1\%$. The 2dFGRS result
$b^2=1.0\pm 0.1$ can then be used at $z=0$;
\begin{equation}
b^2(0)=1.0\pm 0.1.
\end{equation}
We use Eq.~(9) to normalize $E$ in Eq.~(8) for all DE
models.

\begin{figure}[pb]
\centerline{\psfig{file=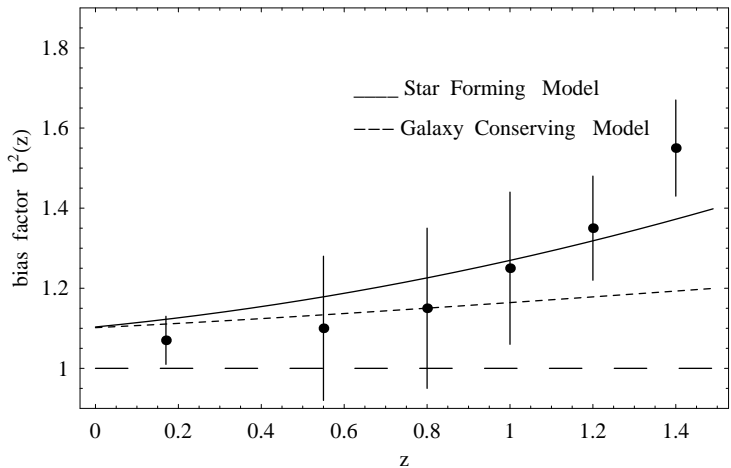,width=10cm}} \vspace*{8pt}
\caption{Analytic models of linear bias evolution, $b^2(z)$ for
the flat $\Lambda$CDM model
($\Omega_M=0.3,\,\Omega_{\Lambda}=0.7$), normalized at
$b_\Lambda^2(z=0)=1$ with a deviation $F=+10\%$, compared with the
$b_\Lambda^2$ for the volume-limited sample data with ${\mathcal
M}^c_{\mathcal B}=-20+5 \log h$ from the first-epoch VIMOS-VLT
Deep Survey (VVDS). The solid and the short-dashed lines
correspond to the star-forming and galaxy conserving models of
$b^2(z)$, respectively. The long-dashed line is $b^2(z)\equiv 1$
for comparison. The bias at $z= 0.17$ (the effective depth of the
2dFGRS survey) was inferred from the 2dFGRS
galaxies,\,\protect\cite{Marinoni} which had the same median
luminosity of the volume-limited VVDS sample, $L/L^\ast\sim 2$.}
\end{figure}

This article is based on the requirement that $b^2(z=0)\cong 1$ to
an accuracy of $10\%$ for all viable DE models, which is
consistent with the results of the above analytic models.  It is
also consistent with the numerical calculations of Somerville et
al., who showed that $b^2_0\simeq 1.0$ for both the CDM and
$\Lambda$CDM models, ($\Omega_M=1.0,\,\Omega_\Lambda =0$) and
($\Omega_M=0.3,\,\Omega_\Lambda =0.7$), respectively.
\par
We can compare the predictions of the above analytic models
against observations of galaxy formation as a function of
redshift. In Fig.~1, we show the galaxy conserving model
(short-dashed curve) and the star forming model (continuous curve)
for $\Lambda$CDM, normalized to $b^2=1.1$ at $z=0$. The curves are
in agreement with the $\Lambda$CDM $b^2(z)$ from the first-epoch
VIMOS-VLT Deep Survey (VVDS) in Ref.\,\refcite{Marinoni} (Marinoni
et al.). (We do not plot the merging model since, as was shown by
Marinoni et al., it does not describe the redshift evolution of
the bias of their galaxies well.)
\par
In Fig.~1, the values of $E$ and $A$ in Eq.~(8) are $E=0.1$ and $A=1.0$ for
the galaxy conserving and star forming models. Using the value of $E$ for the
star-forming model in Eqs.~(4)-(7), we find that the dark matter fluctuations
for the massive $L_\ast$ galaxies,
$\sigma(M)=4.1,$ its present linear extrapolated value. The fact that $\sigma(M)=4.1$ is greater
than $\delta_c=1.69,$ implies that the halo of the $L_\ast$ galaxy has already
collapsed, which is reasonable.

\section{DE and the growth of density fluctuations}

The nature of DE is still unknown and there are many alternative
models which try to explain it. In addition to the popular
cosmological model of a constant vacuum energy, described by a
cosmological constant, there are models that modify gravity as
well as DE those that parametrize the DE EOS, $w(a)= w_0+w_a
(1-a)$, setting values for $w_0$ and $w_a$. For these models, the
Friedmann equation can be written in a general form in terms of an
effective EOS.\,\cite{lindmnras} Modelling the DE as an ideal
fluid in a flat Universe, we can write the Friedmann equation as
\begin{equation}
\frac{H^2(z)}{H_0^2} = \Omega_{M}^0(1+z)^3 + (1-\Omega_{M}^0)\, e^{3\int_0^z
d\ln(1+z')[1+w(z')]}\,, \label{friedeq}
\end{equation}
or
\begin{equation}
\frac{H^2(z)}{H_0^2} = \Omega_{M}^0 (1+z)^3 + \frac{\delta H^2}{H_0^2}\,,
\label{igno}
\end{equation}
where $H_0$ is the present value for the Hubble parameter,
$\Omega_M^0$ is the present normalized matter density, and $\delta
H^2/H_0^2$ depends upon the DE model. The EOS $w(z)$ for
the DE can be written as
\begin{equation}
w(z)\equiv -1+{1\over3}{d\ln\delta H^2/H_0^2\over d\ln(1+z)}\,.\label{effw}
\end{equation}
Linear growth of a density fluctuation, $D= \delta\rho/\rho$,
depends on the EOS. We define the growth factor, $G\equiv D/a$,
where the cosmological scale factor is $a\equiv 1/(1+z)$ and $G$
is normalized to unity at $z\sim 1100$, the recombination
epoch.\,\cite{lindmnras} In terms of $G$, we have
\begin{equation}
G^{\prime\prime}(a)+ \left[\frac{7}{2}-\frac{3}{2}\frac{w(a)}{1+X(a)}
\right]\frac{G^{\prime}(a)}{a}+
\frac{3}{2}\frac{1-w(a)}{1+X(a)}\frac{G(a)}{a^2}=0\,, \label{ggrowth}
\end{equation}
where $X(a)$ is defined as
\begin{equation}
X(a)=\frac{\Omega_M^0 \,a^{-3}}{{\delta H^2}/{H^2_0}}\,.
\end{equation}
For large $X$ we recover the matter dominated behavior $D\sim a$.

It is generally required that the DE was very small
compared to the cold DM for redshifts $z\gtrsim 10$ in
order for the latter to create the structure in the Universe at
redshifts $z \lesssim 10$. In this paper, we analyze only dark
energy models in which the DE density at the
recombination redshift or higher ($z \gtrsim 1100$) was
negligible, as in the $\Lambda$CDM model.
\par
One of the DE models studied was the Vacuum Metamorphosis
Model (VMM), which has zero DE at recombination. A second
model was the Brane World Model (BWM), in which the ratio of the
DE density to the DM density is proportional to
$(1+z)^{-3}$ and negligible at high redshifts. A third model was
the supergravity SUGRA model, which becomes a CDM model at high
redshifts with negligible DE at recombination.
\par
Much effort has been made to obtain the density power spectrum
normalization at the recombination era. For example, Spergel et
al. used the three year WMAP data with a $\Lambda$CDM model to
obtain the density power spectrum normalization $A_S$ at
recombination.\,\cite{spergel}
\par
Our DE model test is independent of $A_S$ since it is
primarily dependent on the evolution of the DM and
radiation densities at $z=1100$, the recombination era, or higher,
when DE was small or negligible. The models that we
examine here have negligible DE for $z\gtrsim 1100$ and,
thus, a negligible effect on $A_S$. It is to be noted that the
$\Lambda$CDM model, used by Spergel et al., is one such model, in
which DE density was one billionth that of the total
energy density at the recombination era.
\par
For our DE model test, it is not necessary to obtain the
exact value of the normalization factor. At a given epoch, the
only relevant difference between the different models studied is
the value of the DE density, all other factors being
equal. Since at the recombination epoch, all of the models we
examined had a negligible DE density, they are
essentially identical and their effect on the normalization factor
is identically negligible. Thus, an uncertainty in the
normalization factor (which is not known to better than $\sim
20\%$) does not enter into the uncertainty of $\sim 10\%$ in the
factor $F$ [Eqs.~(15) and (16)], used in our DE
model test, since all models are normalized at the recombination
era.
\par
It is important to emphasize the difference between the objective
of Spergel et al. and ours. The objective of Spergel et al. was to
obtain a value for $A_S$ as accurate as possible as well as other
parameters, such as the normalized DM density $\Omega_M$,
using one DE model, the $\Lambda$CDM model. Our objective here, is to discard DE models which have
negligible DE densities at the recombination era, but that dominate
the total energy density of the Universe at present.
\par
Galaxies began to form at a redshift $z\sim 20,$ growing until
the present time ($z\sim 0$). The formation and
growth of the galaxies, described by their fluctuations
$\xi_{gg}$, are greater for larger DM fluctuations $\xi_{dm}$ in
the redshift interval $z\sim 20$ to $z\sim 0$. Since $\xi_{dm}$
monotonically increases with a decrease in redshift and DE does
not effect it at very high redshifts ($z\gg 20$), $\xi_{dm}(0)$
is sensitive to the DE model  and the growth factor.
Our test, $b^2(0)\simeq 1,$ does
not tell us whether the growth factor or $\xi_{dm}(0)$ needs to be
big or small. What $b^2(0)\simeq 1$ does say is that whatever
$\xi_{gg}(0)$ is,
$\xi_{dm}(0)$ must accompany it such that $b^2(0)\equiv
\xi_{gg}(0)/\xi_{dm}(0)\simeq 1$.
\par
We define the deviation from the standard $\Lambda$CDM model by
\begin{equation}
F=\left\vert \frac{D^2 -D_{\Lambda}^{\,2}}{D^2}\right\vert\,\Big|_{z=0}\,,
\end{equation}
where $D^2_{\Lambda}$ are the density fluctuations in the standard flat
$\Lambda$CDM model and $D^2$ are the density fluctuations of an
arbitrary DE model. The functions $D/a$ and $D_{\Lambda}/a$ are
normalized to unity at the recombination era, $z\simeq 1100$,
where $a\equiv 1/(1+z)$ is the cosmic scale factor. Therefore $F$ in
Eq.~(15) is concerned only with the relative growth of the density
fluctuations between a given DE model and the flat
$\Lambda$CDM model, from the recombination epoch to the present
era. It thus has the important characteristic of being independent
of the power spectrum normalization at recombination.
\par
The linear bias parameter is defined by  $b^2= P_{gg}/P_{mm}$,
where $P_{gg}$ is a galaxy distribution power spectrum (e.g., that
of 2DFGRS or SDSS) and $P_{mm}(z)\propto
(\delta\rho/\rho)^2=D^2(z)$ is the DM power spectrum.  The
difference in the DE models is in the evolution of
$\delta\rho/\rho$, which occurs after the recombination era at
redshifts $z\lesssim 10$. For this reason, we use the factor $F$, which
examines the effect of DE models on the evolution of
$\delta\rho/\rho$ after the recombination era. The DE
models that we investigate do not effect the evolution of
$\delta\rho/\rho$ before the recombination era and, therefore,
$(\delta\rho/\rho)_{\rm rec}$ is the same for all the DE
models studied. Thus, Eq.~(15) becomes
\begin{equation}
F=\left\vert \frac{b^2 -b_{\Lambda}^{\,2}}{b^2_{\Lambda}
}\right\vert\,\Big|_{z=0}\,.
\end{equation}
A maximum $10\%$ deviation of $b^2$ from $b_{\Lambda}^2$ with $b_{\Lambda}^2\simeq
1.0$ at $z=0$ implies that the maximum $F$ is $F_{\rm max}=0.1$.

\begin{table}[ph]
\tbl{Best fit values of $w_0$ and $w_a$ for the EOS,
$w(a)=w_0+w_a(1-a)$, for the Gold SNIa dataset \protect\cite{comp}
with a deviation $F=0.10 \pm 0.02$ and a matter density
$\Omega_{M}^{0} = 0.28 \pm 0.02$.} {\begin{tabular}{@{}cccc@{}}
\toprule $\,\,\,\,\,\,\,\,\,\,\,\,$&$\,\,\,\,\,\,w_a\,\,\,\,\,\,$
&
$\,\,\,\,\,\,w_0\,\,\,\,\,\,$& $\,\,\,\,\,\,\,\,\,\,\,\,$\\
\colrule
    &$1.53$&$-1.72\pm 0.02$&\\
    &$1.63$&$-1.77\pm 0.02$&\\
    &$1.73$&$-1.82\pm 0.02$&\\
    &$1.83$&$-1.86\pm 0.02$&\\
    &$1.93$&$-1.82-0.01$&\\
    &$2.03$&$-1.89-0.01$&\\
    &$2.9$&$-1.86-0.02$&\\ \botrule
\end{tabular} \label{tablinear}}
\end{table}

\subsection{DE models described by a parametrized EOS}

We first discuss a parametrization for the EOS,
$w(a)=w_0+w_a(1-a)$, which has been widely used for DE models
since it is well-behaved at high redshifts, unlike $w(z)=w_0+w_1
z$, for example, which diverges at high $z$. This parametrization
was introduced by Ref. \refcite{comm1}. The best fit parameters,
$w_0$ and $w_a$, which are consistent with the Gold SNIa dataset,
were found to be in the intervals $-1.91\leqslant w_0\leqslant
-1.25$ and $1.53\leqslant w_a\leqslant 5.05$.\,\cite{comp}
Assuming that $F=0.10\pm 0.02$ and $\Omega_M^0=0.28\pm 0.02$, we
obtain the best fit values, $-1.91\le w_0\le -1.72$ and $1.53\le
w_a\le 2.9$, shown in Table~\ref{tablinear}.

\begin{table}[ph]
\tbl{The deviation $F$ for the BWM and the VMM, respectively, as a
function of the matter density $\Omega_M^0$. The values $H_0 r_c$,
$m^2$ and $z_j$ are defined in $\S$ 3.2.} {\begin{tabular}{@{}c c
c c c c@{}} \toprule
$\Omega_M^0$&$F_{\rm{BWM}}$&$H_0r_c$&$F_{\rm{VMM}}$&$m^2$&$z_j$\\
\colrule
    0.26    &0.27    &1.4     &0.29    &11  &1.4\\
    0.28    &0.27    &1.4     &0.24    &11  &1.4\\
    0.3     &0.26    &1.4     &0.20    &11  &1.3\\
    0.32    &0.25    &1.5     &0.16    &11  &1.2\\
    0.34    &0.25    &1.5     &0.13    &11  &1.2\\
    0.36    &0.24    &1.6     &0.095   &10  &1.1\\
    0.72    &0.11    &3.6     &0.24    &8   &0.5\\ \botrule
\end{tabular} \label{tabbwmvmm}}
\end{table}

\subsection{Brane-world and vacuum metamorphosis models}

In the BWM,\,\cite{bwm} gravity is modified by adding a
five-dimensional Einstein-Hilbert action, that dominates at
distances that are larger than the crossover length $r_c$, which
defines an effective energy density,
$\Omega_{bw}=(1-\Omega_M^0)^2/4 =1/(4H_0^2r_c^2)$, for a flat
Universe. The factor $\delta H^2/H_0^2$ in Eq.~(\ref{igno}) then
becomes
\begin{equation}
\delta H^2/H_0^2 = 2\Omega_{bw}+2\sqrt{\Omega_{bw}}\sqrt{\Omega_M^0(1+z)^3+
\Omega_{bw}}\,. \end{equation}
In the VMM,\,\cite{vmm} the vacuum contributions are due to a
quantized massive scalar field, which is coupled to gravity. For
$z<z_j$, $\delta H^2/H_0^2$ in Eq.~(\ref{igno}) is
\begin{equation}
\delta H^2/H_0^2 = (1-m^2/12)(1+z)^4+m^2/12-\Omega_M^0(1+z)^3\,,
\end{equation}
where $z_j=[m^2/(3 \Omega_M^0)]^{1/3}-1$ and $m^2=3
\Omega_M^0[(4/m^2)-(1/3)]^{-3/4}$. Both the BWM and the VMM can be
described by the EOS, $ w(a)=w_0+w_a(1-a)$,
with $(w_0,w_a)=(-0.78,0.32)$ and $(w_0,w_a)=(-1,-3)$,
respectively.\,\cite{lindastro}

It can be seen from Table~2 that an agreement within  $10\%$
between the VMM and the $\Lambda$CDM model is possible only for a
matter density $\Omega_M^0\approx 0.36$. For the BWM, an agreement
with the $\Lambda$CDM model within  $10\%$ is possible only if the
matter density $\Omega_M^0\approx 0.72$. However, both of these
values for $\Omega_M^0$ are greater than the observed value.


\begin{figure}[pb]
\centerline{\psfig{file=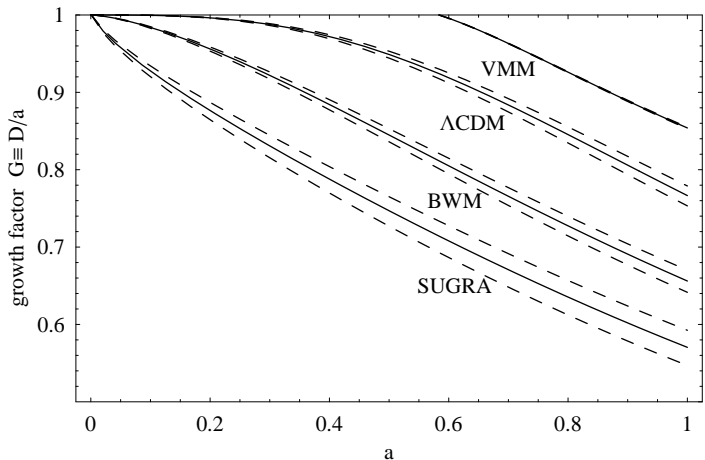,width=10cm}} \vspace*{8pt}
\caption{The growth of density fluctuations,
$G=\left[\left({\delta\rho}/{\rho}\right)/a\right]$, for the VMM,
$\Lambda$CDM, BWM, and the SUGRA models. The dashed lines show the
deviation of the matter density, $\Omega_M^0 =0.28\pm 0.02$.}
\label{growthfig}
\end{figure}
\subsection{Supergravity model}

The SUGRA model is very attractive for explaining the acceleration
of the Universe.\,\cite{sugra} It can be described by the EOS of
$\S$ 3.1, with $w_0=-0.82$ and $w_a=0.58$.\,\cite{lindmnras} This
equation of state is in agreement with observations for the low
redshift SNIa dataset and galaxy distribution
data.\,\cite{Riess,solevi} However, $F$ for this model makes it
inviable.
\par
Figure 2 shows the growth of the density fluctuations
$G=\left[\left({\delta\rho}/{\rho}\right)/a\right]$ as a function of
$a$ for the BWM, the VMM, and the $\Lambda$CDM model. It shows that
the growth of $\delta\rho/\rho$ is smaller for the BWM and SUGRA
model than for the $\Lambda$CDM model. The deviation $F$ for the
SUGRA model is $F_{\rm SUGRA}\approx 0.38_{\,+0.02}^{\,+0.04}$ for
$\Omega_M^0=0.30_{-0.02}^{-0.04}$, which  gives a bias parameter,
normalized to the $b^2_{\Lambda CDM}(z=0)=1$, appreciably greater
than unity: $b^2_{SUGRA}(z=0)=1.38$.

\section{Conclusions and Discussion}

In this paper we suggested a new test for the viability of DE
models, based on the value of the bias parameter $b$, the ratio of
galaxy to DM fluctuations, at $z=0$. If it were the
case that we knew nothing about galaxy formation, $b$ could, in
principle, have been anything at all, i.e., very much less or very much
greater than unity. However, our present knowledge of galaxy
formation from analytic models and numerical simulations,
indicates that $b^2$ is close to unity at $z\simeq 0$. This
information can be used to discard DE models that do not predict
$b$ close to unity at $z=0$.
\par
We studied three popular analytic models for $b(z)$. We showed
that the equations for $b$ in all three analytic models can be
reduced to the form of a known constant plus the term
$E[D(z=0)/D(z)]^{\alpha}$, where $\alpha=1$ or 2 and $E$ is a free
parameter. Using the value for $b$ obtained by the 2dFGRS
consortium\,\cite{cole05} for all three models predict
$b^2(0)=1.0\pm 0.1$ for all DE models. This value of $b$ is also
in agreement with numerical simulations that evaluated $b^2(0)$
for the $\Lambda$CDM and CDM ($\Lambda=0$)
models.\,\cite{somerville99} Since this value of $b^2(0)$ is
indicated by numerical simulations as well as by all three popular
analytic models, which are normalized by the 2dFGRS consortium
result for the $\Lambda$CDM model, we suggest the condition that
$b^2(0)=1\pm 0.1$ at $z=0$ as a new test for the viability of DE
models.
\par
Obtaining $b(0)$ from galaxy observations involves a complex
process of combining data from all types of galaxies (see e.g.,
Ref.\,\refcite{swanson}). These complexities reflect the galaxy
formation process. As in the standard analysis for the evaluation
of the bias parameter $b$ (see e.g., Cole et al. Ref.
\refcite{cole05}), we assume that $b$ is independent of scale for
$k\leqslant 0.1 h\,{\rm Mpc}^{-1}$ and that all galaxies are
normalized to a standard massive bright galaxy with luminosity
$L_{\ast}$. Cole et al. made the following normalization of the
bias parameter to a galaxy of luminosity $L_{\ast}$ for the
galaxies in the 2dF Galaxy Redshift Survey:
\begin{itemize}
\item {Luminosity Normalization: $b(L)=0.85+0.15(L/L_{\ast});$}
\item {Red Galaxy Subset Normalization: $b {\rm{(red)}}=1.3[0.85+0.15(L/L_{\ast})];$} and
\item {Blue Galaxy Subset Normalization: $b {\rm{(blue)}}=0.9[0.85+0.15(L/L_{\ast})].$}
\end{itemize}
\par
Obtaining $b(0)$ from numerical simulations of galaxy formation is
also not simple. It is difficult to build models for galaxy
populations of dark halos that can robustly relate the amplitude
of large-scale galaxy clustering at better than the $10\%$
level.\,\cite{gao07} This, in part, is the reason that we have a
$10\%$ limit on the accuracy of our test, which examines whether
$b=1$ at $z=0$ for a viable DE model.
\par
The numerical simulations of the Virgo Consortium are consistent
with our test for DE models. Very high-resolution simulations have
been made by the Virgo Consortium for the $\Lambda$CDM
model.\,\cite{springel05} They found that for galaxies with $M_B<
-17$ at $z=0$ on the largest scales, the galaxy power spectrum has
the same shape as that of the DM, but with a slightly lower
amplitude, corresponding to a bias $b=0.92$. Samples of brighter
galaxies have a bias close to $b\simeq 1.0$.
\par
In this article, we related $b(0)$ to the $\Lambda$CDM value of
$b$, $b_\Lambda(0)$, using a function $F$. The bias parameter $b$
is related to the factor $F$ by $F=|(b^2/b_\Lambda^2)-1|$ at $z=0$
(Eq.~16). Thus, a maximum deviation of $10\%$ of $b^2(z=0)$ from
$b_\Lambda^2(z=0)$ with $b_\Lambda^2(z=0)\simeq 1.0$ implies a
$\sim 10\%$ deviation of $F$ from zero or $F_{\rm max}=0.1$. We
investigated DE models that make a negligible contribution to the
total energy density before the recombination era and, thus, a
negligible contribution to the density power spectrum
normalization factor $A_S$ at the recombination era. The function
$F$ has the important characteristic of being independent of
$A_S$. We calculated the value of $F$ numerically for several
well-known DE models from the growth equation for
$\delta\rho/\rho$. The constraints from the Gold SNIa
data\,\cite{comp} and the condition that $F=0.1$ restrict the
values of the parameters of the linear EOS, $w(a)=w_0+w_a(1-a)$,
for dark energy. It was found that the best fit values of $w_0$
and $w_a$ are $-1.86<w_0<-1.72$, with $1.53<w_a<2.9$. For $z\sim
0.5 - 1$, where $w(a)$ is sensitive to the supernova data, and
$w_0\sim -1$.
\par
The BWM and VMM were then studied using the factor $F$. We showed
that these DE models do not satisfy our DE model
test, with $F=0.1$ and $\Omega_M^0=0.28\pm 0.02$. The BWM has
$F=0.26$ and the VMM, $F=0.20$. Thus BWM and VMM are not viable DE
models.
\par
Finally, we analyzed the SUGRA model for the above parametrized
EOS with $w_0=-0.82$ and $w_a=0.58$. $F$ was found to be very
large: $F_{\rm SUGRA}\approx 0.38_{\,-0.02}^{\,+0.04}$ for
$\Omega_M^0=0.30_{-0.02}^{+0.04}$, giving $b^2_{SUGRA}(z\cong
0)=1.38$, which is appreciably greater than unity. Thus the SUGRA
model is also not a viable DE model.



\section*{Acknowledgments}

R.O. thanks the Brazilian agencies FAPESP (00/06770-2) and CNPq
(300414/82-0) for partial support. A.P. thanks FAPESP (grants
03/04516-0 and 00/06770-2), MECYT/FEDER (project
2004-04582-C02-01) and DURSI Generalitat de  Catalunya (project
2005SGR00564). A.P. also thanks the Dept. ECM of the Univ. of
Barcelona for their hospitality.



\end{document}